\documentclass[11pt, a4paper]{article}
\pdfoutput=1

\usepackage[usenames, dvipsnames]{color}
\usepackage[T1]{fontenc}
\usepackage[latin1]{inputenc}
\usepackage{lmodern}
\usepackage{amsmath}
\usepackage[pdftex]{graphicx}
\usepackage{lastpage}
\usepackage{amssymb} 
\usepackage{hyperref}
\usepackage{cite}
\usepackage{soul}

\usepackage{subcaption}

\usepackage{esvect}
\renewcommand*\vec{\vv}

\usepackage{xspace}
\newcommand*{\ie}{i.e.\@\xspace}
\newcommand*{\eg}{e.g.\@\xspace}
\newcommand*{\fig}{Fig.\@\xspace}
\newcommand*{\rhs}{r.h.s.\@\xspace}
\newcommand*{\lhs}{l.h.s.\@\xspace}

\newcommand{\ex}{\text{e}}
   
\usepackage{slashed}
\usepackage{braket}
\usepackage{simplewick}
\usepackage{bbm}

\newcommand\numberthis{\addtocounter{equation}{1}\tag{\theequation}}

\usepackage{xifthen}
\newcommand*\diff{\mathrm{d}} 
\newcommand*\ldiff[2][]{ \ifthenelse{\isempty{#1}}{ \diff #2}{\diff^#1#2} \,} 
\let\limitint\int 
\renewcommand{\int}{\limitint \!} 

\DeclareMathOperator{\Tr}{Tr}

\usepackage{subcaption}

\usepackage{titling}
\usepackage{authblk}

\title{Infrared Divergences and Quantum Coherence}

\author[1,2]{C\'{e}sar G\'{o}mez\thanks{cesar.gomez@uam.es}}
\author[1]{Raoul Letschka\thanks{raoul.letschka@csic.es}}
\author[2,3]{Sebastian Zell\thanks{sebastian.zell@campus.lmu.de}}
\affil[1]{Instituto de F\'{i}sica Te\'orica UAM-CSIC, Universidad Aut\'onoma de Madrid, \mbox{Cantoblanco, 28049 Madrid, Spain}}
\affil[2]{Arnold Sommerfeld Center, Ludwig-Maximilians-Universit\"at, \mbox{Theresienstra\ss e 37, 80333 M\"unchen, Germany}}
\affil[3]{Max-Planck-Institut f\"ur Physik, F\"ohringer Ring 6, 80805 M\"unchen, Germany}

\newcommand*\B{B_{\alpha,\, \beta}} 
\newcommand*\Bp{B_{\alpha,\, \beta'}}
\newcommand*\Bb{B_{\beta,\, \beta'}}
\newcommand*\BbDiag{B_{\beta,\, \beta}}
\renewcommand*\S[1][]{ \ifthenelse{\isempty{#1}}{R_{\alpha,\, \beta}}{R_{\alpha,\, \beta #1}}}
\newcommand*\Sp[1][]{ \ifthenelse{\isempty{#1}}{R_{\alpha,\, \beta'}}{R_{\alpha,\, \beta'#1}}}
\newcommand*\Sb[1][]{ \ifthenelse{\isempty{#1}}{R_{\beta,\, \beta'}}{R_{\beta,\, \beta'#1}}}
\newcommand*\G{\Gamma_{\alpha,\, \beta}}
\newcommand*\I{\mathcal{I}(\bra{\beta}R\ket{\beta'})}
\newcommand*\Il{\mathcal{I}^{\text{loop}}(\bra{\beta}R\ket{\beta'})}
\newcommand*\If{\mathcal{I}^{\text{full}}(\bra{\beta}R\ket{\beta'})}
\newcommand*\F{\mathcal{F}_{\alpha,\, \beta}(\gamma)}
\newcommand*\Rt{\rho^{(\alpha),\, \text{dec}}_{\beta \beta'}}
\newcommand*\Rtni{\rho^{(\alpha),\, \text{dec}}}
\newcommand*\R{\rho^{(\alpha),\, \text{coh}}_{\beta \beta'}}
\newcommand*\Rni{\rho^{(\alpha),\, \text{coh}}}
\newcommand*\RDiag{\rho^{(\alpha),\, \text{coh}}_{\beta \beta}}

\begin{document}

\allowdisplaybreaks

\maketitle

\vspace{\baselineskip}
\begin{abstract}
In theories with long-range forces like QED or perturbative gravity, only rates that include emitted soft radiation are non-vanishing. Independently of detector resolution, finite observables can only be obtained after integrating over the IR-component of this radiation. This integration can lead to some loss of quantum coherence. In this note, however, we argue that it should in general not lead to full decoherence. Based on unitarity, we suggest a way to define non-vanishing off-diagonal pieces of the IR-finite density matrix. For this IR-finite density matrix, we estimate the dependence of the loss of quantum coherence, \ie of its purity, on the scattering kinematics. 
\end{abstract}

\newpage

\tableofcontents

\section{Introduction and Summary}
The definition of an operative infrared-finite $S$-matrix for theories like QED or perturbative gravity is a task yet to be accomplished. The problem arises from the fact that the amplitude for a given process $(\alpha \rightarrow \beta)$ without any emission of soft IR-radiation vanishes because of soft radiative loops. These contributions from soft loops can be removed by taking into account the IR-divergent soft emission processes $(\alpha \rightarrow \beta + \text{\it IR-radiation})$ \cite{BN, YFS, weinberg}, where IR-radiation is defined as the radiation that is emitted from external lines. The cancellation does not take place at the level of the $S$-matrix, but only for physical observables such as the differential rate. There we integrate in an inclusive way over soft IR-emissions in order to extract from the soft emission amplitudes the divergent term canceling the contribution due to radiative loops. It is very important to stress that this inclusive integration over IR-radiation is not due to any practical limitation on resolution, that in reality obviously exists. Instead, it is needed to compensate for the problem created by the radiative loops. The IR-finite quantities obtained, following this standard recipe, depend on an IR-energy scale that sets the amount of IR-radiated energy as well as the upper bound on the energy of individual radiated quanta.

 There is an alternative way to deal with IR-divergences, namely the coherent state approach suggested in \cite{FK}, in which initial and final states are dressed with a coherent state of soft IR-radiation. In this way, the integration over soft IR-emission is implicitly included in the dressing factors.  However, whereas energy conservation is automatically implemented in the definition of the integration measure of soft IR-emission, this is not the case for the coherent state dressing. Therefore, the key difficulty with this approach is to implement energy conservation as well as to describe the dependence on the IR-energy scale. 

 As we have reviewed, we can define IR-finite rates by integrating over soft IR-emission. A very interesting question raised recently in \cite{semenoff, semenoff2} is if we can go one step further and define an IR-finite density matrix. Obviously, its diagonal is determined by the known IR-finite rates. So the task consists in determining  the IR-limit of the off-diagonal pieces of this density matrix. These elements contain the information about how much quantum coherence we lose by tracing over the soft IR-radiation. This is an important question because this tracing is, as stressed above, not due to any limit on detector resolution but -- at least in the present formalism -- a prerequisite for IR-finiteness.

The result of the calculation in \cite{semenoff, semenoff2} was that almost all off-diagonal elements are zero in an arbitrary scattering process. This finding is surprising since it was obtained in the absence of any environment and does not depend on detector resolution. So it seems to imply that the requirement of IR-finiteness alone inevitably leads to an almost completely decohered density matrix. If this were true, we would find a very disturbing physical picture. Then the final state of an arbitrary scattering process would be a fully decohered mixed state and it would become impossible to have an IR-finite description of quantum interference phenomena. Concretely, we can perform a double-slit experiment with the products of a scattering event, \ie we take the final state of scattering as initial state for the double-slit experiment. If the final scattering state were really fully decohered, it would never be able to lead to interference patterns in flagrant conflict with experimental evidence. 

As is well-known, decoherence is an omnipresent phenomenon in any experimental setup because of inevitable interaction with the environment. Nevertheless, this fact does not preclude quantum interference phenomena. The reason is that the experiment takes place in a time smaller than the time of decoherence.  Even in the absence of an environment, it is conceivable that integration over IR-radiation leads to decoherence depending on its entanglement with the hard scattering data.
 A natural estimate of the resulting decoherence time would be that it scales inversely with the energy in IR-modes.  Therefore, if one waits for an {\it infinite} amount of time, an arbitrarily small energy in IR-modes could lead to full  decoherence \cite{semenoff}.

In contrast, our aim is to propose a density matrix that can describe the quantum coherence observed in experiments, \ie that has non-vanishing off-diagonal elements. In doing so, we will achieve two goals. First, we point out why the off-diagonal elements vanish. The reason is a soft loop contribution, very similar to the one discussed above that leads to vanishing amplitudes. Its origin lies in the contribution of zero-energy IR-radiation. Secondly, we show that the requirement of IR-finiteness alone does not inevitably lead to full decoherence by providing an explicit example of a density matrix that is IR-finite but not fully decohered.

We will construct this density matrix using the optical theorem. If the only IR-finite quantity were the rate, the optical theorem would only be informative about forward scattering. But if we are interested in  obtaining non-trivial information about the imaginary part for generic amplitudes, we can solely achieve this with non-zero off-diagonal pieces of the density matrix. For generic states, non-trivial information about unitarity of the $S$-matrix can therefore only be obtained with IR-finite off-diagonal pieces of the density matrix. In what follows, we shall suggest a concrete definition of these off-diagonal pieces based on an IR-finite version of the optical theorem. 

An important comment to be made at this point is that the so defined IR-finite and non-vanishing off-diagonal elements of the density matrix do not necessarily give rise to complete purification. After all, we are tracing over soft IR-radiation in order to achieve IR-finiteness and this tracing, although needed by finiteness, can lead to some quantum decoherence. This decoherence, as already mentioned, is due to the entanglement between the soft IR-radiation we integrate over and the rest of scattering products. One origin of this entanglement obviously is energy conservation.
As we shall show, energy conservation leads to a dependence of the IR-finite density matrix on the IR-kinematical factors, which scale with the coupling and are only sensitive to the initial and final scattering state. Thus, the von Neumann entropy of the IR-finite density matrix depends on these kinematical factors allowing us to study how IR-decoherence depends on the scattering kinematics. 

So far, we have only discussed tracing over soft IR-radiation, which is required for IR-finiteness. In an experimental setup, there is obviously a second reason for tracing, namely a finite detector resolution. In that case, one traces over {\it all} soft radiation, \ie also over non-IR radiation. Since this non-IR radiation is entangled with the scattering products through the well-known damping forces and depends on the details of the scattering process, it is clear that decoherence occurs in that case.  In this note, however, we will not consider the effect of a finite detector resolution but solely focus on IR-finiteness.

\section{Review of Treatment of IR-Divergences}
\subsection{Finite Rates}
\label{ssec:finiteRate}
 
We consider the transition from an initial state $\ket{\alpha}$ to a final state $\ket{\beta}$, which is described by a generic tree level amplitude $\S^0$. Here $\S^0$ is the nontrivial part of the $S$-matrix, $S_{\alpha,\, \beta} = \delta_{\alpha,\, \beta} + i \S$, and does not contain soft loops. If one takes into account soft loops in this process, their effect can be resummed and exponentiated so that we obtain \cite{YFS, weinberg}:
\begin{equation}
	\S^{\text{loop}} = \left(\frac{\lambda}{\Lambda}\right)^{\B/2} \S^0 \,,
	\label{softLoop} 
\end{equation}
where $\lambda$ is an IR-cutoff and $\Lambda > \lambda$ a UV-cutoff for the loop integration.  The exponent $\B$, which we shall discuss in a moment, is a non-negative number which depends on the kinematical data of the process $\alpha \rightarrow \beta$. For $\B\neq0$, this IR-correction leads to a vanishing amplitude in the limit $\lambda \rightarrow 0$.

 In the case of QED, the exponent $\B$ is given by
\begin{equation}\label{BQED}
\B= -\frac{1}{8 \pi^2} \sum_{\substack{n\in \alpha, \, \beta\\ m\in \alpha, \, \beta}} \eta_n \eta_m e_n e_m \beta_{nm}^{-1} \ln \left(\frac{1+\beta_{nm}}{1-\beta_{nm}}\right) \,,
\end{equation}
where both sums run over all external particles. Here $e_n$ marks the electric charge of a particle, $\eta_n=1$ for an outgoing particle and $\eta_n=-1$ for an ingoing one and $\beta_{nm}$ is the relative velocity:
\begin{equation}
	\beta_{nm} = \left(1-\frac{m_n^2 m_m^2}{(p_n \cdot p_m)^2}\right)^{1/2} \,.
\end{equation}
As usual, $m_n$ is the mass and $p_n$ the 4-momentum. In the case of gravity, we have\footnote
{The massless limit of this expression is finite and was derived in \cite{veneziano}.}
\begin{equation}
\B= \frac{G}{2 \pi} \sum_{\substack{n\in \alpha, \, \beta\\ m\in \alpha, \, \beta}} \eta_n \eta_m m_n m_m \frac{1+\beta_{nm}^2}{\beta_{nm}(1-\beta_{nm}^2)^{1/2}} \ln \left(\frac{1+\beta_{nm}}{1-\beta_{nm}}\right) \,,
\end{equation}
where $G$ is the gravitational constant. Both in gravity and QED, $\B$ is suppressed by the coupling constant. So it can only get big in a regime of strong coupling.\footnote
{For example, in the case of scattering of two gravitons of ultra-planckian center of mass energy $s$ into an  arbitrary number of final gravitons, we have $\B=4 \,c\, G s /\pi$, where $0\leq c \leq \ln 2$ \cite{veneziano}. The lower bound is reached if all final gravitons are collinear with the initial ones and the upper bound corresponds to all final gravitons orthogonal to the initial ones.}
 In \cite{semenoff}, the interesting statement was proven that $\B=0$ if and only if the outgoing current in $\ket{\beta}$ matches the ingoing current in $\ket{\alpha}$ antipodally at each angle. Thus, the contribution of soft loops leads to a vanishing amplitude for $\lambda \rightarrow 0$ unless we are in a very special situation. 

The standard approach to deal with the problem of the vanishing IR-correction is to include the effect of soft IR-emission, \ie to consider a different process $\alpha \rightarrow \beta + \gamma$, where at most an energy $\epsilon$ is contained in radiation $\gamma$.  According to the soft factorization theorems, the resulting amplitude is 
\begin{equation}
 \S[\gamma] = \F \S \,.
 \label{softFactorization}
\end{equation}
The factor $\F$ depends only on the initial and final state and on the additional IR-radiation. In the case of QED, it reads
\begin{equation}
	\F = \sum_{n\in \alpha, \, \beta} \sum_{\pm} \sum_{i \in \gamma} \frac{e_n \eta_n }{ \sqrt{(2\pi)^{3} 2|\vec{k}_i|}} \frac{p_n^\mu \varepsilon_{\mu, \, \pm}^\star(\vec{k}_i)}{p_n \cdot k_i- i \eta_n \epsilon} \,,
\end{equation}
where $k_i$ is the 4-momentum of a soft photon and $\vec{k}_i$ its momentum. Moreover $\varepsilon_\mu$ is its polarization vector and $\pm$ labels the helicity. In gravity, we obtain analogously
\begin{equation}
\F = \sum_{n\in \alpha, \, \beta} \sum_{\pm} \sum_{i \in \gamma} \frac{\sqrt{8 \pi G} \eta_n }{\sqrt{(2\pi)^{3}2|\vec{k}_i|}} \frac{p_n^\mu p_n^\nu\varepsilon_{\mu\nu, \, \pm}^\star(\vec{k}_i)}{p_n \cdot k_i- i \eta_n \epsilon} \,.
\end{equation}
Tracing over the radiation $\gamma$ yields the rate
\begin{equation}
	\G^{\text{full}} := \sum_{\gamma}  |\S[\gamma]^\text{loop}|^2 =  \left(\frac{\epsilon}{\lambda}\right)^{\B} f(\B)  |\S^\text{loop}|^2 =   \left(\frac{\epsilon}{\Lambda}\right)^{\B} f(\B) |\S^0|^2 \,.
	\label{finiteRate}
\end{equation}
The additional factor $f(\B)$ is due to energy conservation and reads \cite{YFS}
\begin{equation}
f(x) = \frac{\ex^{-\gamma x}}{\Gamma(1+x)} \,,
\end{equation}
where $\gamma$ is Euler's constant and $\Gamma$ is the gamma function. For small $x$, it can be approximated as
\begin{equation}
f(x) = 1 - \frac{\pi^2}{12}x^2 \,.
\label{smallF}
\end{equation}
For large $x$, it scales as
\begin{equation}
f(x) \sim \frac{1}{x!} \,.
\label{largeF}
\end{equation}
Combining the effects of soft loops and soft emission, one obtains a rate which is independent of the IR-cutoff and in particular finite for $\lambda \rightarrow 0$. 

\begin{figure}
	\centering 
	\begin{subfigure}{0.6\textwidth}
			\centering
		\includegraphics[height=8.5\baselineskip]{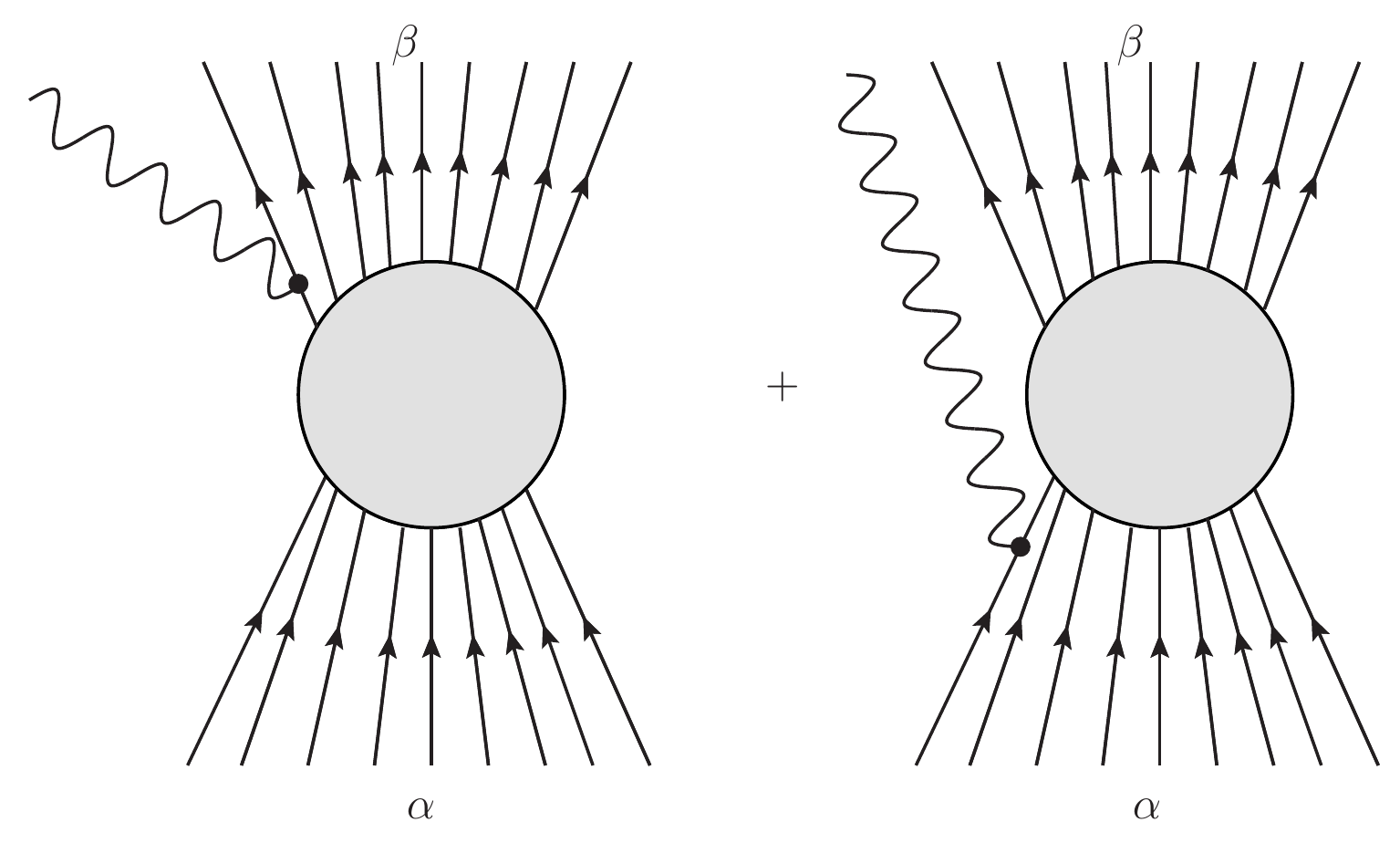}
		\caption{ IR-radiation }
		\label{fig:External_lines}
	\end{subfigure}
	\begin{subfigure}{0.35\textwidth}
			\centering
		\includegraphics[height=8.5\baselineskip]{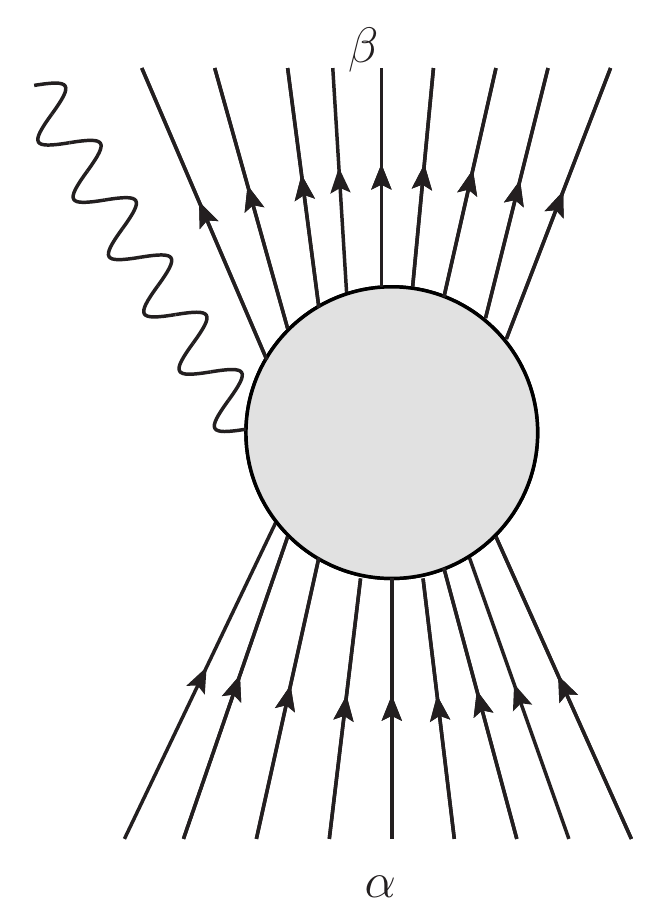}
		\caption{Non-IR radiation}
		\label{fig:Internal_lines}
	\end{subfigure}
	\caption{ Comparison of IR-radiation and non-IR radiation in the scattering process $\alpha \rightarrow \beta$. IR-radiation is the IR-divergent part of emission from external lines, \ie from an incoming or an outgoing particle. In contrast, non-IR radiation comes from emission from internal lines and is IR-finite.}
\end{figure}

It is important to remark that $\gamma$ does not include all kinds of radiation with energies below $\epsilon$ but only {\it IR-radiation}, which is defined as the part of radiation which leads to a divergent amplitude for $\lambda \rightarrow 0$. Diagrammatically, IR-radiation is due to emission from external legs (see \fig \ref{fig:External_lines}), whereas soft emission from internal lines is infrared-finite (see \fig \ref{fig:Internal_lines}), \ie non-IR. We shall elaborate on this point shortly.

\subsubsection*{The Meaning of IR-Symmetries}
 As a side note, we remark how the discussion of IR-divergences could shed light on the recently discussed {\it soft symmetries}. In references \cite{stromingerBMSInvariance, stromingerMasslessQED, mohd, campiglia,stromingerQED, stromingerRevisited, mirbabayi, stromingerLecture, campiglia2} it has been suggested that in theories with long range forces there exist an infinite set of charges $Q_{\epsilon}$ that can be associated with the asymptotic states of scattering processes. If we consider a process $\alpha \rightarrow \beta$, these charges relate the current in $\beta$ to the current in $\alpha$ at each angle, namely the incoming current should match the outgoing current antipodally. However, this is precisely the condition under which the corresponding kinematical soft factor $\B$ vanishes, \ie we observe that
\begin{equation} \label{softSymmetries}
Q_{\epsilon}\ket{\alpha} = Q_{\epsilon}\ket{\beta} \quad\iff \quad \B=0 \,.
\end{equation}
Thus, imposing soft symmetries is equivalent to restricting to the special class of processes that are IR-finite even without including soft IR-emission.\footnote
{Only the soft charge of zero modes is automatically conserved: Since those are decoupled, a zero mode in the initial state always implies a zero modes in the final state at the antipodal angle. We refer the reader to \cite{porrati1, sever, raoul,mischa, porrati2,porrati3} for recent discussions on decoupling of zero-energy modes.}
The scattering processes that satisfy this constraint form a set of zero measure.
 As we have reviewed, however, {\it all} scattering processes are IR-finite after including the emission of soft IR-radiation. So in general, there is no physical reason to restrict to final states that satisfy the constraint \eqref{softSymmetries}.

\subsection{Finite Density Matrix}
\label{ssec:finiteDensity}
As a next step, we investigate the density matrix  which results from the initial state $\ket{\alpha}$ after computing the effect of soft loops and tracing over soft IR-radiation $\gamma$. Generalization of the above procedure gives
\begin{equation}
\Rtni
 = \sum_{\beta \beta' \gamma} \S[\gamma]^{\text{loop}} \Sp[\gamma]^{\text{loop}\,*} \ket{\beta}\bra{\beta'} \,.
	\label{decoheredDensityDefinition}
\end{equation}
We can use the soft factorization theorem \eqref{softFactorization} to evaluate the matrix element of the density matrix \cite{semenoff}:
\begin{equation}
\Rt = \S^0 \Sp^{0\,*} 	 \left(\frac{\lambda}{\epsilon}\right)^{\Bb/2}  \left(\frac{\epsilon}{\Lambda}\right)^{(\B+\Bp)/2}\!\! f\left(\frac{\B+\Bp-\Bb}{2}\right) \,.
\label{decoheredDensity}
\end{equation}
 Even after tracing over soft IR-emission, a non-negative power of $\lambda$ survives. Its exponent is the kinematical factor $\Bb$ of the hypothetical scattering process $\beta \rightarrow \beta'$. This means that $\lambda$ vanishes if and only if the currents in $\beta$ and $\beta'$ match  antipodally at each angle. On the diagonal, this is the case so that we obtain
\begin{equation}
\rho^{(\alpha),\, \text{dec}}_{\beta \beta}= |\S^0|^2  \left(\frac{\epsilon}{\Lambda}\right)^{\B} f(\B) \,.
\end{equation}
As it should, the diagonal gives the known rate \eqref{finiteRate}. For generic states $\beta$ and $\beta'$, in which the currents do not match angle-wise, however, a positive power of $\lambda$ survives on the off-diagonal. Thus, the corresponding off-diagonal elements vanish in the limit $\lambda \rightarrow 0$. In a generic case, the resulting density matrix $\Rtni$ therefore is mostly decohered, thereby justifying its superscript. This finding is independent of the specific process and the coupling strength.

\subsubsection*{Decoherence and Zero-Energy Modes}
From the previous discussion it is easy to identify the root of the former decoherence of the density matrix. 
 Since almost all off-diagonal elements vanish for $\lambda \rightarrow 0$, it follows from the form \eqref{decoheredDensityDefinition} of the density matrix that in this limit 
\begin{equation}\label{maximalEntanglement}
\sum_\gamma \S[\gamma]^{\text{loop}}\Sp[\gamma]^{\text{loop}\, *} \sim \delta_{\beta, \, \beta'} \,.
\end{equation}
Thus, full decoherence after tracing over IR-radiation is equivalent to maximal entanglement between the hard state $\ket{\beta}$ and the IR-radiation.
  However, this behavior only occurs in the limit $\lambda \rightarrow 0$, independently of the values of the other scales $\epsilon$ and $\Lambda$. This shows that {\it only radiated quanta of zero energy are responsible for the decoherence} and immediately raises the question, whether the decoherence derived above really corresponds to a physical effect. Namely, the actual decoupling of the zero energy modes should lead to recovering quantum coherence, at least partially.

\subsubsection*{Effect of Non-IR Radiation}
As a final remark, we want to point out that not all off-diagonal elements vanish. In particular, this happens if $\beta$ and $\beta'$ have the same current and only differ in their soft non-IR radiation. Moreover, since photons are uncharged, $\beta$ and $\beta'$ also yield a non-vanishing off-diagonal element in QED when they differ by hard photons. In a process in which a lot of radiation is produced, a sizable amount of off-diagonal elements therefore survives. For example, in the process of electron-positron annihilation in QED, all final states have the same electronic content (namely none) so that no decoherence at all takes place. In general, of course, taking into account soft non-IR radiation does not suffice to obtain an (approximately) pure density matrix since, as said above, the vanishing of all off-diagonal elements whose currents do not match angle-wise leads to a significant amount of decoherence. In particular, this is true if we consider a weakly coupled process, in which final states without hard radiation dominate. Our goal is to find a procedure that leads to an (approximately) pure final state also for those.

\section{Proposal for IR-Finite Density Matrix with \mbox{Coherence}}

\subsection{Modified Density Matrix from Optical Theorem}
Thus, we have to change the procedure explained in section \ref{ssec:finiteDensity} so that the resulting density matrix is -- at least approximately -- pure. This means that we have to modify the off-diagonal elements without changing the diagonal. This will be achieved in three steps. First, we will show that the optical theorem relates the imaginary part of the amplitude for the process $\beta \rightarrow \beta'$ with the elements $\rho^{(\alpha)}_{\beta\beta'}$ of the density matrix discussed above.
Secondly, we will derive an IR-finite version of the optical theorem. As third step, this IR-finite optical theorem will enable us to define a density matrix that possesses IR-finite off-diagonal elements.

For generic states $\ket{\beta}$ and $\ket{\beta'}$, the optical theorem reads 
\begin{equation}
-i\left(\bra{\beta}R\ket{\beta'} - \bra{\beta'}R\ket{\beta}^\star\right) = \sum_{\sigma} \bra{\beta} R \ket{\sigma} \bra{\beta'}R\ket{\sigma}^\star \,,
\label{opticalTheorem}
\end{equation}
where the states $\ket{\sigma}$ form a complete set and we used that $S=1+iR$.  In terms of the matrix elements $\rho^{(\sigma)}_{\beta\beta'}$ of the density matrix for the process $\sigma\rightarrow\beta$, one can write \eqref{opticalTheorem} as:
\begin{equation}
\I = \sum_{\sigma} \rho^{(\sigma)}_{\beta\beta'} \,,
\label{opticalTheorem2}
\end{equation}
 where we introduced the abbreviation 
\begin{equation}
\I := -i\left(\bra{\beta}R\ket{\beta'} - \bra{\beta'}R\ket{\beta}^\star\right) \,.
\label{definitionI}
\end{equation}
Now we study the effect of soft modes in the optical theorem, \ie we split the Hilbert space in IR-radiation $\gamma$ and all other states $\alpha$:
\begin{equation}
\I = \sum_{\alpha}\sum_{\gamma} \bra{\beta}R\ket{\alpha, \gamma} \bra{\beta'} R \ket{\alpha, \gamma}^\star \,.
\end{equation}
We can use that the contributions of IR-radiation factorize. Since we have $\B = B_{\beta,\, \alpha}$ and moreover all soft correction factors are real, we get
\begin{equation}
\I  = \sum_{\alpha} \Rt \,,
\end{equation}
where the matrix elements $\Rt$ of the decohered density matrix \eqref{decoheredDensityDefinition} appear. Plugging in the result \eqref{decoheredDensity}, we obtain
\begin{align*}
\I &= \left(\frac{\lambda}{\epsilon}\right)^{\Bb/2}\\
 &\sum_{\alpha} \S^0 \Sp^{0\,*} 	   \left(\frac{\epsilon}{\Lambda}\right)^{(\B+\Bp)/2} f\left(\frac{\B+\Bp-\Bb}{2}\right) \,.
\numberthis \label{decoheredOptical}
\end{align*}
It is crucial here that $\Bb$ does not depend on $\alpha$. This expression vanishes in the limit $\lambda \rightarrow 0$. However, this does not come as a surprise. By including IR-radiation in the sum over all intermediate states, we effectively introduced soft loops in the  process $\beta \rightarrow \beta'$. This is the reason why we obtain the factor $(\lambda/\epsilon)^{\Bb/2}$, which comes from including soft loops of energies below $\epsilon$ in the process $\beta \rightarrow \beta'$. So in the above computation, we should replace $\I \rightarrow \Il$.

\begin{figure}[h]
	\centering 
	\begin{subfigure}{0.3\textwidth}
		\centering
		\includegraphics[height=17\baselineskip]{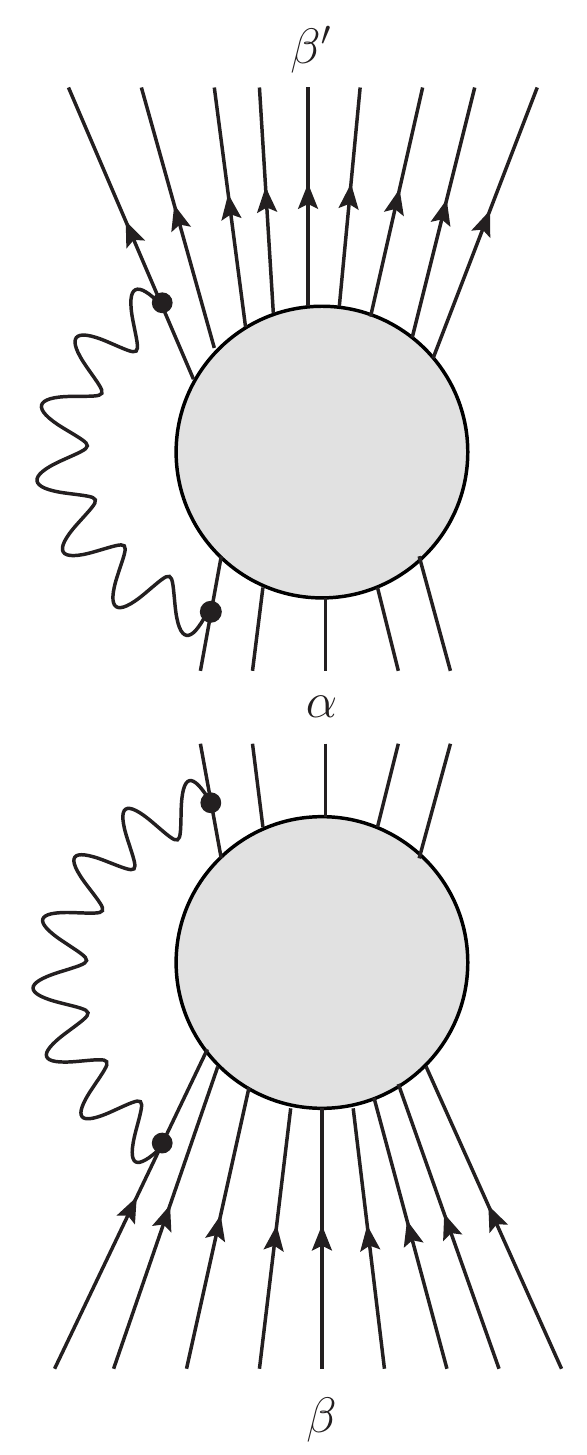}
		\caption{  $(\lambda/\Lambda)^{(\B+\Bp)/2}$  }
		\label{fig:densityMatrixLoop}
	\end{subfigure}
	\begin{subfigure}{0.3\textwidth}
		\centering
		\includegraphics[height=17\baselineskip]{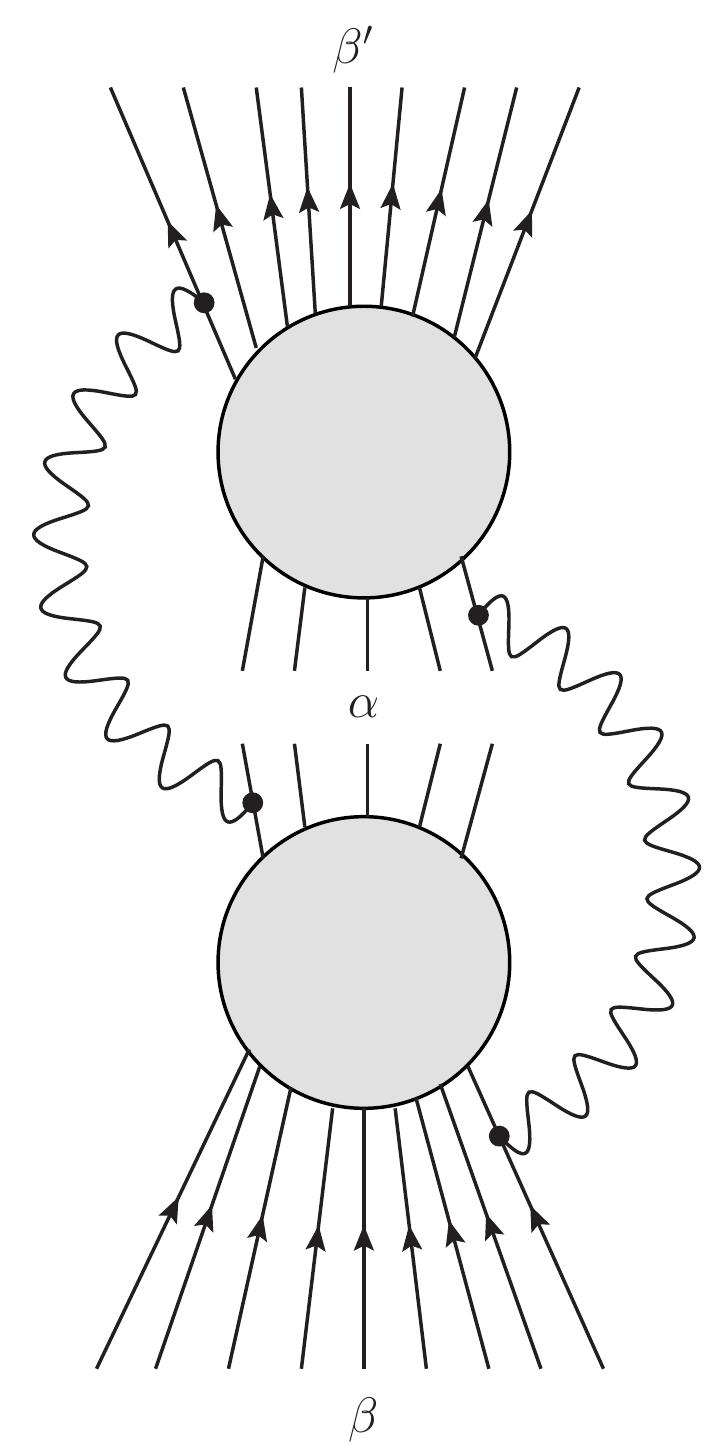}
		\caption{ $(\epsilon/\lambda)^{(\B+\Bp)/2}$ }
		\label{fig:densityMatrixEmission1}
	\end{subfigure}
	\begin{subfigure}{0.3\textwidth}
		\centering
		\includegraphics[height=17\baselineskip]{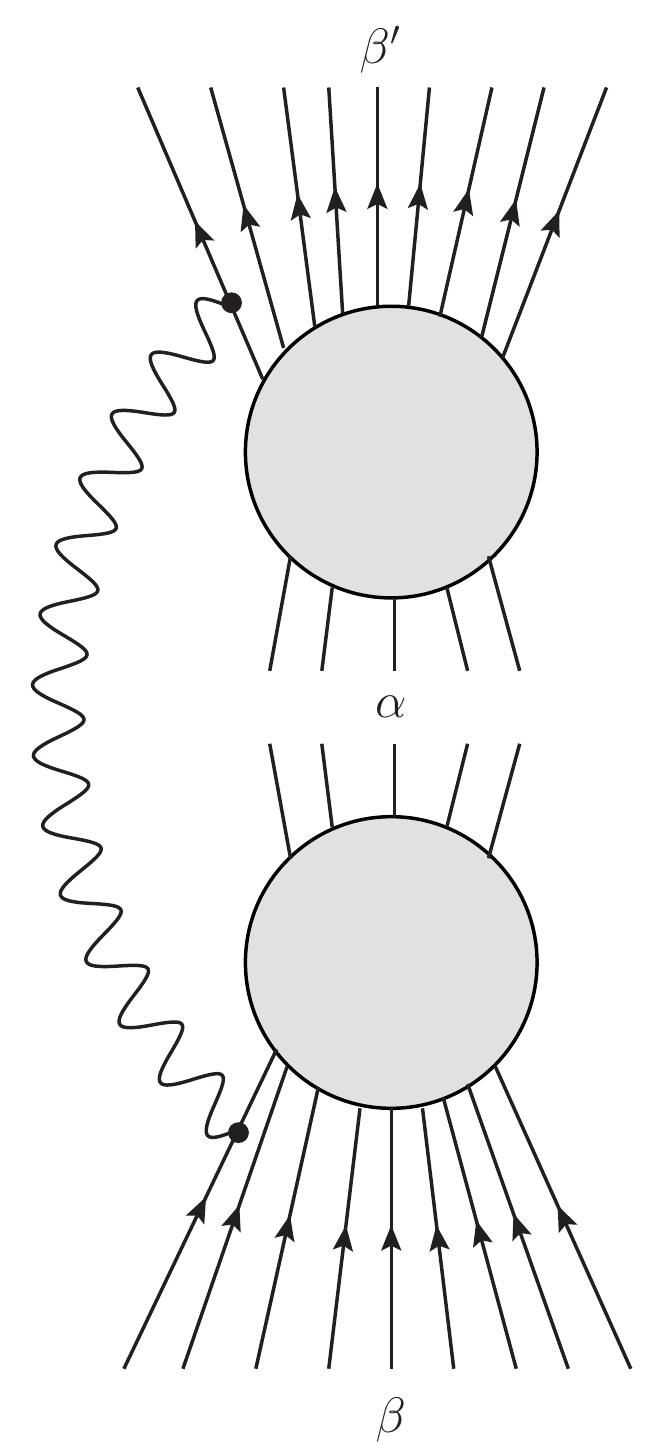}
		\caption{ $(\lambda/\epsilon)^{\Bb/2}$}
		\label{fig:densityMatrixEmission2}
	\end{subfigure}
	\caption{Diagrammatic representation of the different contributions to the density matrix element $\Rt$. The first contribution is due to soft loops. The second and the third one come from tracing over emitted IR-radiation. The product of the first two contributions gives an IR-finite result, but because of the third one, most off-diagonal elements vanish for $\lambda \rightarrow 0$. The reason is that by tracing over IR-emission  in the processes $\alpha \rightarrow \beta$ and $\alpha \rightarrow \beta'$, we effectively introduce a soft loop from the perspective of the process $\beta \rightarrow \beta'$. }
	\label{fig:densityMatrix}
\end{figure}

 That the density matrix $\Rt$ appears in the optical theorem also gives us the opportunity to better understand where its different contributions come from, as is illustrated in \fig \ref{fig:densityMatrix}. In particular, we can identified the origin of the factor $(\lambda/\epsilon)^{\Bb/2}$, which leads to vanishing off-diagonal elements in the density matrix and therefore to decoherence:  As already said, including soft IR-emission in the processes $\alpha \rightarrow \beta$ and $\alpha \rightarrow \beta'$ effectively generates soft loops in the process $\beta \rightarrow \beta'$ (see \fig \ref{fig:densityMatrixEmission2}). 
Those loops lead to a vanishing matrix element $\Rt$ unless the currents in $\beta$ and $\beta'$ match antipodally  at each angle.

The second step is to derive an IR-finite version of the optical theorem. We remark that this is an important question on its own. Namely, if no IR-finite version existed for a generic process $\beta \rightarrow \beta'$, the optical theorem would become meaningless in a gapless theory except for the case $\beta \rightarrow \beta$ of forward scattering. Since we have concluded that soft loops are the reason why relation \eqref{decoheredOptical} is zero in the limit $\lambda \rightarrow 0$, it is clear how to obtain a non-trivial finite answer. Namely, we have to include soft IR-emission in the process $\beta \rightarrow \beta'$. This means that we proceed in full analogy to \eqref{finiteRate} and define
\begin{equation}
\big|\If\big|^2 := \sum_{\gamma} 	\big|\mathcal{I}^{\text{loop}}(\bra{\beta}R\ket{\beta',\gamma})\big|^2 \,.
\end{equation} 
Plugging in the definition \eqref{definitionI} of $\Il$, we can generalize the standard computation displayed in \eqref{finiteRate} to obtain the simple result
\begin{equation}
\big|\If\big|^2 = \left(\frac{\epsilon}{\lambda}\right)^{\Bb} f(\Bb)	\, \big|\Il\big|^2 \,.
\end{equation} 
Since the soft factors that appeared due to the inclusion of the emission of IR-modes are real and positive, there is a natural definition of the square root of the above equation that preserves all analytic properties of $\Il$:
\begin{equation}
	\If = \left(\frac{\epsilon}{\lambda}\right)^{\Bb/2} \sqrt{f(\Bb)}	\, \Il \,.
\end{equation}
Plugging in the explicit form \eqref{decoheredOptical} of $\Il$, we finally obtain
\begin{align*}
\If &= \sqrt{f(\Bb)} \\
&\sum_{\alpha} \S^0 \Sp^{0\,*} 	   \left(\frac{\epsilon}{\Lambda}\right)^{(\B+\Bp)/2} f\left(\frac{\B+\Bp-\Bb}{2}\right) \,. \numberthis 
\label{finiteOpticalTheorem}
\end{align*} 
This is the result of applying the standard recipe \cite{BN, YFS, weinberg} for dealing with infrared divergences to the optical theorem. While we believe that the definition \eqref{finiteOpticalTheorem} can in general give an IR-finite meaning to the optical theorem, the only important point for the present work is that we can derive an IR-finite density matrix from it.

Consequently, the third step is to use the IR-finite version \eqref{finiteOpticalTheorem} of the optical theorem to obtain a density matrix that is not fully decohered. We do so in an effective description in which IR-modes are fully integrated out. Namely, we define the density matrix as the one that has to appear on the \rhs of the optical theorem \eqref{opticalTheorem2} when full IR-finite quantities are used on the \lhs:
\begin{equation}
\If = \sum_{\alpha} \R \,,
\label{opticalTheoremFull}
\end{equation}
where the sum over $\alpha$ no longer contains IR-modes. Thus, we obtain the full IR-finite density matrix as
\begin{equation}
	\R := \S^0 \Sp^{0\,*}  \left(\frac{\epsilon}{\Lambda}\right)^{(\B+\Bp)/2}\!\! \sqrt{f(\Bb)} f\left(\frac{\B+\Bp-\Bb}{2}\right) \,.
	\label{coheredDensity}
\end{equation}
 In comparison with the decohered density matrix \eqref{decoheredDensity}, there are two changes: The factor $(\lambda/\epsilon)^{\Bb/2}$ was removed and the function $\sqrt{f(\Bb)}$ was added. {The first change is crucial since it alone suffices to avoid full decoherence.} It is important to note that this form of the density matrix solely follows from requiring that the IR-finite description in terms of $\If$ and $\Rni$ is unitary in the sense that it fulfills the optical theorem in the form \eqref{opticalTheoremFull}, which we obtained after integrating out IR-modes.

 As a side note, we remark how one can understand our approach in the diagrammatic representation in terms of \fig \ref{fig:densityMatrix}. When we sum over all possible intermediate states in the optical theorem, we also include those where soft IR-quanta are emitted or absorbed by the hard modes defining the intermediate state $\ket{\alpha}$, as is displayed in \fig \ref{fig:densityMatrixEmission1}. Additionally, however, there is the contribution of \fig \ref{fig:densityMatrixEmission2}, in which we sum over IR-radiation that is emitted from $\beta$ and then absorbed in $\beta'$. It is fully insensitive to the intermediate state $\ket{\alpha}$ and the one that leads to the factor $(\lambda/\epsilon)^{\Bb/2}$, which is responsible for decoherence.  Our recipe provides us with a concrete way to avoid decoherence by removing this contribution.

In summary, the logical flow of our approach can be described as follows. We consider a given scattering process $\beta \rightarrow \beta'$,  whose amplitude is zero due to IR-divergences. But if we IR-regulate the amplitude, \ie do not take $\lambda \rightarrow 0$, its imaginary part is still non-trivial if we have branch cuts reflecting the threshold of inelastic processes. Since the existence of these cuts is not affected by adding IR-soft radiation, it should survive in the IR-limit. In other words, in the same way that we know that the actual scattering $\beta \rightarrow \beta'$ is non-trivial once we add, in an appropriate way, soft IR-radiation, we expect that the inelastic part of the scattering is equally non-vanishing after including soft IR-radiation. What we have presented is a simple recipe to derive from this physics picture a natural characterization of  quantum decoherence.

\subsection{Resulting Entropy}
We proceed to discuss the modified density matrix \eqref{coheredDensity}.  First, we note that the diagonal is the same as for the decohered density matrix \eqref{decoheredDensity}: 
\begin{equation}
\RDiag = |\S^0|^2  \left(\frac{\epsilon}{\Lambda}\right)^{\B}  f(\B) \,,
\label{coheredDensityDiag}
\end{equation}
which follows from $\BbDiag=0$. Thus, we obtain the known rate \eqref{finiteRate}. As it should be, our modification of the density matrix does not change the rates.
The important question is how pure the density matrix \eqref{coheredDensity} is. As a first step, we investigate what off-diagonal elements would be required to obtain a completely pure result. From the diagonal elements \eqref{coheredDensityDiag} it follows that we would need
\begin{equation}
	\left(\R\right)^{\text{pure}} = \S^0 \Sp^{0\,*}  \left(\frac{\epsilon}{\Lambda}\right)^{\B/2+\Bp/2}  \sqrt{f(\B)}\sqrt{f(\Bp)} \,.
\end{equation}
In that case, the density matrix would be pure since we could write it as
\begin{equation}
\left(\Rni\right)^{\text{pure}} = \ket{\Psi}\bra{\Psi} \,,
\end{equation}
where 
\begin{equation}
\ket{\Psi} = \sum_\beta \S^0   \left(\frac{\epsilon}{\Lambda}\right)^{\B/2} \sqrt{f(\B)}\ket{\beta} \,.
\end{equation}
Thus, only the functions $f(B)$, which arise due to energy conservation, lead to decoherence. We can parametrize the deviation from purity as the quotient of the factor in the full modified density matrix \eqref{coheredDensity} and the factor required for purity:
\begin{equation}
c^{(\alpha)}_{\beta,\, \beta'}=	\frac{\sqrt{f(\Bb)} f(\B/2+\Bp/2-\Bb/2)}{\sqrt{f(\B)f(\Bp)}} \,.
\end{equation}
 So the deviations of the $c^{(\alpha)}_{\beta,\, \beta'}$ from $1$ determine the decoherence and full coherence corresponds to $c^{(\alpha)}_{\beta,\, \beta'}=1$. 

 In order to study decoherence in more detail, we will restrict ourselves to the regime of weak coupling. In that case, all functions $f(B)$ are small, \ie we can use the approximation \eqref{smallF}. Then we get to leading order:
\begin{equation}
c^{(\alpha)}_{\beta,\, \beta'} = 1+  \frac{\pi^2}{48}\left((\B - \Bp)^2 + 2 \Bb(\B + \Bp) - 3\Bb^2\right) \,.
\end{equation}
  This shows that all $c^{(\alpha)}_{\beta, \beta'}$ are arbitrarily close to $1$ for sufficiently weak coupling,  \ie decoherence can be avoided by decreasing the coupling. We will make this statement quantitative, \ie we determine an upper bound on the decoherence that can arise. We will estimate in terms of the von Neumann entropy $S=-\Tr \Rni \ln \Rni$. If all off-diagonal element were zero, the maximal entropy would be given by $S=\ln d_H$, where $d_H$ is the dimension of the hard Hilbert space. This maximal entropy would be reached if all final hard states were equally probable, \ie all diagonal elements were equal. For our estimate, we will therefore restrict ourselves to a density matrix in which all diagonal elements are equal.  Such a density matrix is pure if all elements, \ie also the off-diagonal ones, are equal. In order to derive the upper bound on the entropy, we can consequently define $\Delta_{\text{max}}:= \max_{\beta,\, \beta'}  |1-c^{(\alpha)}_{\beta,\, \beta'}|$ and then multiply the off-diagonal elements of the pure density matrix, in which all entries are equal, by the function $c := 1-\Delta_{\text{max}}$.\footnote
{At this point, one can wonder why we could not use $c := 1+\Delta_{\text{max}}$ instead. The reason is that any $c>1$ would lead to an unphysical density matrix with negative eigenvalues. Note that is it nonetheless not excluded that some $c^{(\alpha)}_{\beta,\, \beta'}$ are bigger than $1$.}
 In this setup, the eigenvalues of the density matrix are\footnote
 {These are the eigenvalues of a quadratic matrix of dimension $d_H$ that has $1/d_H$ on the diagonal and $c/d_H$ on all off-diagonal elements.  A linearly independent set of eigenvectors $v_i$ is given by the entries $(v_1)_k = 1$ and $(v_i)_k = \delta_{k1} - \delta_{ki}$ for $i\in [2,\, d_H]$.}
\begin{equation}
	e_1 = \frac{1+(d_H-1)c}{d_H} \quad \text{and}\ e_i = \frac{1-c}{d_H} \ \text{for}\ i\in [2,\, d_H] \,.
\end{equation} 
To leading order in $1-c$, this gives the bound 
\begin{equation}
	S < (1-c) \ln\left(\frac{d_H}{1-c}\right) \,.
\end{equation}
 As expected, we obtain $S=0$, \ie purity, for $c=1$. Full decoherence can only be obtained  in the limit $c=0$.  This confirms that the entropy is always small if the coupling is weak enough. So at least in the case of weak coupling, our formalism is able to describe the interference phenomena that we observe experimentally.

Clearly, our estimate no longer works in the regime of strong coupling. However, from this fact it does not follow that a sizable amount of decoherence has to occur in that case. In particular, the fact that the hard amplitudes depend strongly on the final state in the strong coupling regime could prevent the generation of entropy. It would be interesting to investigate this question in a concrete setup, \eg the process of $2\rightarrow N$-scattering proposed in \cite{2toN}, whose infrared behavior was already studied in \cite{veneziano}.

\section{Outlook}
In order to describe interference phenomena that we observe in Nature, one needs an IR-finite density matrix with non-vanishing off-diagonal elements. In this note, we have constructed such an IR-finite density matrix by supplementing the usual scattering calculation with additional input from an IR-finite version of the optical theorem. In this way, we have obtained an effective description of interference phenomena that allowed us to study how coherence depends on scattering kinematics. However, an important task for the future \cite{future} will be to derive an IR-finite density matrix from a first-principle $S$-matrix calculation alone and to understand how the density matrix that we have constructed here can be embedded in such a framework.  

Moreover, it would be important to investigate how the von Neumann entropy due to IR-tracing grows for strong coupling. In the case of gravity, this corresponds to scattering at ultra-planckian energies, where we can expect black holes to form. So black holes occur precisely in the regime in which decoherence could become significant. Therefore, it is natural to ask what role infrared physics could play in the process of black hole formation and evaporation. In particular, based on earlier work \cite{HPS1, HPS2}, the interesting question was raised in \cite{stromingerNew} if IR-modes alone could suffice to purify Hawking radiation. This would mean that tracing over IR-modes is the main reason why Hawking radiation is mixed.

Our results show, however, that IR-finiteness alone does not necessarily lead to decoherence. So there is a priori no reason to expect that IR-modes lead to decoherence and conversely, that resolving them would lead to purification. Moreover, as already argued in \cite{sebastian}, they cannot lead to any significant amount of decoherence if one only considers the isolated process of evaporation. The reason is that this process is weakly coupled since Hawking radiation gets softer for bigger black holes. So far, however, we have only discussed soft IR-radiation. Thus, it could be interesting to investigate if it is possible to understand purification of evaporation in terms of soft non-IR modes, which do not only depend on initial and final states, but also on the details of the scattering process. The last-mentioned fact makes them a much more involved, but also more promising subject of study.

 \section*{Acknowledgements}
It is a pleasure to thank Gia Dvali for interesting discussions. The work of C.G. was supported in part by Humboldt Foundation and by Grants: FPA 2009-07908 and ERC Advanced Grant 339169 "Selfcompletion". The work of R.L. was supported by the ERC Advanced Grant 339169 "Selfcompletion''.

\providecommand{\href}[2]{#2}\begingroup\raggedright\endgroup

\end{document}